\pdfoutput=1

\RequirePackage[l2tabu,orthodox]{nag}

\documentclass[11pt,letterpaper,reqno,oneside]{article}

\DeclareTextFontCommand{\emph}{\slshape}

\makeatletter
\renewcommand{\paragraph}{%
	\@startsection{paragraph}{4}%
	{\z@}{1.75ex \@plus 1ex \@minus .2ex}{-0.7em}%
	{\normalfont\normalsize\bfseries}%
}
\makeatother

\usepackage[T1]{fontenc}
\usepackage{lmodern}
\usepackage[utf8]{inputenc}

\usepackage[american,german,british]{babel}

\usepackage[activate={true,nocompatibility}]{microtype}

\usepackage{csquotes}

\usepackage[backend=biber,autolang=other,style=apa,maxcitenames=5,sortcites=false]{biblatex}
\DeclareLanguageMapping{british}{british-apa}
\DeclareLanguageMapping{american}{american-apa}
\usepackage{amsmath}
\usepackage{amssymb}
\usepackage{amsfonts}
\usepackage{amsthm}
\usepackage{mathtools}
\usepackage{stmaryrd}

\let\originalleft\left
\let\originalright\right
\renewcommand{\left}{\mathopen{}\mathclose\bgroup\originalleft}
\renewcommand{\right}{\aftergroup\egroup\originalright}

\usepackage{graphicx}
\usepackage{tikz,pgfplots}
\pgfplotsset{compat=1.10}
\usetikzlibrary{shapes.multipart}
\usetikzlibrary{decorations.markings}

\usepackage[title]{appendix}

\usepackage{datetime}
\newdateformat{datestyle}{ {\THEDAY} \monthname[\THEMONTH] \THEYEAR }

\usepackage{enumerate}
\usepackage{enumitem}
\setlist[enumerate,1]{label=(\arabic*)}
\setlist[itemize,1]{label=--}
\setlist[itemize,2]{label=--}
\setlist[itemize,3]{label=--}
\setlist[itemize,4]{label=--}

\usepackage{xcolor}
\usepackage{verbatim}
\usepackage{gensymb}
\usepackage{rotating}
\usepackage{subcaption}
\usepackage{floatpag}
\floatpagestyle{plain}
\usepackage{marvosym}

\usepackage{hyphenat}
\theoremstyle{definition}

\newtheorem{theorem}{Theorem}
\newtheorem{corollary}{Corollary}
\newtheorem{definition}{Definition}

\newtheoremstyle{named}
	{\topsep}					
	{\topsep}					
	{}							
	{0pt}						
	{\bfseries}					
	{}							
	{5pt plus 1pt minus 1pt}	
	{\thmnote{#3}}				
\theoremstyle{named}
\newtheorem{namedthm}{}

\usepackage{xpatch}
\xpatchcmd{\proof}{\itshape}{\proofheadfont}{}{}
\newcommand{\proofheadfont}{\slshape}

\usepackage{hyperref}
\hypersetup{pdfborder=0 0 0}

\usepackage[nameinlink]{cleveref}
\crefname{page}{p.}{pp.}
\crefname{equation}{equation}{equations}
\crefname{section}{section}{sections}
\crefname{subsection}{section}{sections}
\crefname{subsubsection}{section}{sections}
\crefname{appsec}{appendix}{appendices}
\crefname{supplsec}{supplemental appendix}{supplemental appendices}
\crefname{footnote}{footnote}{footnotes}
\crefname{figure}{figure}{figures}
\crefname{table}{table}{tables}
\crefname{theorem}{theorem}{theorems}
\crefname{proposition}{proposition}{propositions}
\crefname{lemma}{lemma}{lemmata}
\crefname{corollary}{corollary}{corollaries}
\crefname{remark}{remark}{remarks}
\crefname{observation}{observation}{observations}
\crefname{example}{example}{examples}
\crefname{fact}{fact}{facts}
\crefname{definition}{definition}{definitions}
\crefname{assumption}{assumption}{assumptions}
\crefname{exercise}{exercise}{exercises}
\crefname{notation}{notation}{notation}
\crefname{claim}{claim}{claims}
\crefname{conjecture}{conjecture}{conjectures}

\newcommand{\dd}{\mathrm{d}}

\newcommand{\E}{\mathbf{E}}

\newcommand{\R}{\mathbf{R}}

\newcommand*{\xslant}[2][76]{%
	\begingroup
	\sbox0{#2}%
	\pgfmathsetlengthmacro\wdslant{\the\wd0 + cos(#1)*\the\wd0}%
	\leavevmode
	\hbox to \wdslant{\hss
		\tikz[
			baseline=(X.base),
			inner sep=0pt,
			transform canvas={xslant=cos(#1)},
		] \node (X) {\usebox0};%
		\hss
		\vrule width 0pt height\ht0 depth\dp0 %
	}%
	\endgroup
}
\makeatletter
\newcommand*{\xslantmath}{}
\def\xslantmath#1#{%
	\@xslantmath{#1}%
}
\newcommand*{\@xslantmath}[2]{%
	\ensuremath{%
		\mathpalette{\@@xslantmath{#1}}{#2}%
	}%
}
\newcommand*{\@@xslantmath}[3]{%
	\xslant#1{$#2#3\m@th$}%
}
\makeatother

\makeatletter
\def\namedlabel#1#2{\begingroup
	#2%
	\def\@currentlabel{#2}%
	\phantomsection\label{#1}\endgroup
}
\makeatother

\makeatletter
\let\save@mathaccent\mathaccent
\newcommand*\if@single[3]{%
	\setbox0\hbox{${\mathaccent"0362{#1}}^H$}%
	\setbox2\hbox{${\mathaccent"0362{\kern0pt#1}}^H$}%
	\ifdim\ht0=\ht2 #3\else #2\fi
	}
\newcommand*\rel@kern[1]{\kern#1\dimexpr\macc@kerna}
\newcommand*\widebar[1]{\@ifnextchar^{{\wide@bar{#1}{0}}}{\wide@bar{#1}{1}}}
\newcommand*\wide@bar[2]{\if@single{#1}{\wide@bar@{#1}{#2}{1}}{\wide@bar@{#1}{#2}{2}}}
\newcommand*\wide@bar@[3]{%
	\begingroup
	\def\mathaccent##1##2{%
	  \let\mathaccent\save@mathaccent
	  \if#32 \let\macc@nucleus\first@char \fi
	  \setbox\z@\hbox{$\macc@style{\macc@nucleus}_{}$}%
	  \setbox\tw@\hbox{$\macc@style{\macc@nucleus}{}_{}$}%
	  \dimen@\wd\tw@
	  \advance\dimen@-\wd\z@
	  \divide\dimen@ 3
	  \@tempdima\wd\tw@
	  \advance\@tempdima-\scriptspace
	  \divide\@tempdima 10
	  \advance\dimen@-\@tempdima
	  \ifdim\dimen@>\z@ \dimen@0pt\fi
	  \rel@kern{0.6}\kern-\dimen@
	  \if#31
	    \overline{\rel@kern{-0.6}\kern\dimen@\macc@nucleus\rel@kern{0.4}\kern\dimen@}%
	    \advance\dimen@0.4\dimexpr\macc@kerna
	    \let\final@kern#2%
	    \ifdim\dimen@<\z@ \let\final@kern1\fi
	    \if\final@kern1 \kern-\dimen@\fi
	  \else
	    \overline{\rel@kern{-0.6}\kern\dimen@#1}%
	  \fi
	}%
	\macc@depth\@ne
	\let\math@bgroup\@empty \let\math@egroup\macc@set@skewchar
	\mathsurround\z@ \frozen@everymath{\mathgroup\macc@group\relax}%
	\macc@set@skewchar\relax
	\let\mathaccentV\macc@nested@a
	\if#31
	  \macc@nested@a\relax111{#1}%
	\else
	  \def\gobble@till@marker##1\endmarker{}%
	  \futurelet\first@char\gobble@till@marker#1\endmarker
	  \ifcat\noexpand\first@char A\else
	    \def\first@char{}%
	  \fi
	  \macc@nested@a\relax111{\first@char}%
	\fi
	\endgroup
}
\makeatother

\usepackage{moreenum}

\newcommand{\mpc}[1]{\text{MPC}(#1)}

\title{\scshape Not Strassen's theorem\thanks{I am grateful to César Barilla, Gregorio Curello and (especially) Ian Jewitt for stimulating conversations on this and related topics. Some sentences in this note are drawn from one of my other papers \parencite{wp}, with my coauthor's kind permission.}}

\author{Ludvig Sinander \\
University of Oxford}

\date{11 November 2026}

\makeatletter
	\AtBeginDocument{ \hypersetup{
		pdftitle = {Not Strassen's theorem},
		pdfauthor = {Ludvig Sinander}
		} }
\makeatother

\begin{document}

\maketitle

\begin{abstract}
	Suppose there is an uncertain state of the world, drawn from a CDF $G$. It is well known that the random posterior mean induced (via Bayes's rule) by a signal must have a distribution $F$ that is a mean-preserving contraction of $G$, and conversely that any mean-preserving contraction is induced by some signal. This is \emph{not} Strassen's (\citeyear{Strassen1965}) theorem! It is a distinct result, versions of which were stated (or at least alluded to) by \textcite{HardyLittlewoodPolya1929}, \textcite{Blackwell1951} and \textcite{Kolotilin2014}\nocite{Kolotilin2018}. I give a precise statement and a complete proof.
\end{abstract}

\section{Setup}
\label{sec:setup}

Throughout this note, we consider random variables taking values in a fixed bounded interval, without loss the unit interval $[0,1]$.%
	\footnote{Explicitly: given a probability space $\mathcal{P} = (\Omega,\mathcal{F},\mu)$, a \emph{$[0,1]$-valued random variable} on $\mathcal{P}$ is an $\mathcal{F}$/$\mathcal{B}$-measurable function $\Omega \to [0,1]$, where $\mathcal{B}$ denotes the Borel $\sigma$-algebra on $[0,1]$.}
We shall use the term `distribution' throughout to denote CDFs concentrated on $[0,1]$.%
	\footnote{That is, increasing and right-continuous functions $F : [0,1] \to [0,1]$ satisfying $F(1)=1$.}

\begin{definition}
	\label{definition:rich}
	Let $\mathcal{P} = (\Omega,\mathcal{F},\mu)$ be a probability space, and let $Y$ be a $[0,1]$-valued random variable on $\mathcal{P}$. We say that $\mathcal{P}$ is \emph{rich relative to $Y$} if and only if there exists a random variable on $\mathcal{P}$ which is statistically independent of $Y$ and uniformly distributed.
\end{definition}

This is a rather mild richness condition. If $Y$ is defined on a probability space $\mathcal{P} = (\Omega,\mathcal{F},\mu)$, then we can easily enlarge the probability space to make it rich relative to $Y$: let $\mathcal{P}' \coloneqq (\Omega \times [0,1],\mathcal{F} \times \mathcal{B},\mu \times \lambda)$, where $\mathcal{B}$ is the Borel $\sigma$-algebra on $[0,1]$ and $\lambda$ is the Lebesgue measure on $[0,1]$, and let $Y'$ be the random variable on $\mathcal{P}'$ given by $Y'(\omega,x) \coloneqq Y(\omega)$ for all $\omega \in \Omega$ and $x \in [0,1]$. Obviously $\mathcal{P}'$ is rich relative to $Y'$, and $Y'$ is `the same' random variable as $Y$ (since it does not vary with $x$).

For two distributions $F$ and $G$,
recall that $F$ is a mean-preserving contraction of $G$ if and only if
\begin{equation*}
	\int_0^x F
	\leq \int_0^x G
	\quad \text{for every $x \in [0,1]$, with equality at $x=1$,}
\end{equation*}
or equivalently if and only if $\int \phi \dd F \leq \int \phi \dd G$
for every convex $\phi : [0,1] \to \R$.%
	\footnote{See e.g. \textcite[§3.A]{ShakedShanthikumar2007}. Another equivalent condition is $\int \phi \dd F \leq \int \phi \dd G$ for every convex \emph{and continuous} $\phi : [0,1] \to \R$.}
We write `$\mpc{G}$' for the set of all mean-preserving contractions of $G$.

\section{The two theorems}
\label{sec:theorems}

We are interested in the following two theorems.

\begin{namedthm}[Strassen's theorem {\normalfont \parencite{Strassen1965}}.]
	\label{theorem:strassen}
	Fix a distribution $G$. Then, for a distribution $F$, the following are equivalent:

	\begin{enumerate}[label=(\alph*)]
	
		\item \label{strassen:exists} There exist a probability space $\mathcal{P}$ and $[0,1]$-valued random variables $X$ and $Y$ on $\mathcal{P}$ such that $X \sim F$, $Y \sim G$, and $\E(Y|X)=X$ a.s.

		\item \label{strassen:mpc} $F$ belongs to $\mpc{G}$.
	
	\end{enumerate}
\end{namedthm}

\begin{theorem}
	\label{theorem:kbhlp}
	Fix a probability space $\mathcal{P}$ and a $[0,1]$-valued random variable $Y$ on $\mathcal{P}$. Assume that $\mathcal{P}$ is rich relative to $Y$, and let $G$ denote the distribution of $Y$. Then, for a distribution $F$, the following are equivalent:

	\begin{enumerate}[label=(\greek*)]
	
		\item \label{kbhlp:exists} There exists a $[0,1]$-valued random variable $X$ on $\mathcal{P}$ such that $X \sim F$ and $\E(Y|X)=X$ a.s.

		\item \label{kbhlp:mpc} $F$ belongs to $\mpc{G}$.
	
	\end{enumerate}
\end{theorem}

I am not sure whom to credit for \Cref{theorem:kbhlp}. It is stated (more or less) in \textcite{Kolotilin2014}\nocite{Kolotilin2018}, though without a complete proof. Versions of the result appear already in \textcite{HardyLittlewoodPolya1929} and \textcite{Blackwell1951}. It seems useful to state it here, and to give a complete proof (below).

The two theorems are different. In \hyperref[theorem:strassen]{Strassen's theorem}, a \emph{distribution} $G$ is given, and we ask for which distributions $F$ we can find (a probability space and) \emph{two} random variables, $X$ and $Y$, such that $(X,Y)$ is a martingale with $X \sim F$ and $Y \sim G$. In \Cref{theorem:kbhlp}, (a probability space and) a \emph{random variable} $Y$ is given, and we ask for which distributions $F$ we can find \emph{one} random variable, $X$, such that $(X,Y)$ is a martingale and $X \sim F$. (It turns out that for the second question, in which a random variable $Y$ (not only its distribution) is given, the answer depends on $Y$ only through its distribution. But that is not a priori obvious.)

To be more precise, the trivial direction of \hyperref[theorem:strassen]{Strassen's theorem} (`\ref{strassen:exists} implies \ref{strassen:mpc}') is identical to the trivial direction of \Cref{theorem:kbhlp} (`\ref{kbhlp:exists} implies \ref{kbhlp:mpc}'),%
	\footnote{Modulo richness, which in fact plays no role. Here is the trivial proof: if we are given a probability space $\mathcal{P}$ and random variables $X$ and $Y$ on $\mathcal{P}$ whose distributions are $F$ and $G$ and which satisfy $\E(Y|X)=X$ a.s., then for any convex $\phi : [0,1] \to \R$, we have $\int \phi \dd F = \E(\phi(X)) = \E(\phi(\E(Y|X))) \leq \E(\E(\phi(Y)|X)) = \E(\phi(Y)) = \int \phi \dd G$ by Jensen's inequality and the law of iterated expectations.}
and the non-trivial direction of \Cref{theorem:kbhlp} immediately implies the non-trivial direction of \hyperref[theorem:strassen]{Strassen's theorem}.%
	\footnote{If we are given distributions $F$ and $G$ such that $F \in \mpc{G}$, then arbitrarily choose a probability space $\mathcal{P}$ and a $[0,1]$-valued random variable $Y$ on $\mathcal{P}$ such that $\mathcal{P}$ is rich relative to $Y$ and $Y \sim G$; then by \Cref{theorem:kbhlp}, there exists a $[0,1]$-valued random variable $X$ on $\mathcal{P}$ such that $X \sim F$ and $\E(Y|X)=X$ a.s.}
By contrast, the non-trivial direction of \Cref{theorem:kbhlp} does not immediately follow from the non-trivial direction of \hyperref[theorem:strassen]{Strassen's theorem}.

It \emph{is} possible to devise a proof of the non-trivial direction of \Cref{theorem:kbhlp} whose first step invokes (the non-trivial direction of) \hyperref[theorem:strassen]{Strassen's theorem}, but at least one further step is required. This extra step can be taken using Kallenberg's transfer theorem \parencite[Theorem~6.10]{Kallenberg2002}, as follows. (Thanks to Ian Jewitt for introducing me to this theorem.)

\begin{proof}[Proof that \ref{kbhlp:mpc} implies \ref{kbhlp:exists}]
	Suppose we are given a probability space $\mathcal{P}$, a $[0,1]$-valued random variable $Y$ on $\mathcal{P}$ whose distribution is denoted $G$, and a distribution $F$, where $\mathcal{P}$ is rich relative to $Y$ and $F$ belongs to $\mpc{G}$. By \hyperref[theorem:strassen]{Strassen's theorem}, there exists a(nother) probability space $\mathcal{P}'$ and $[0,1]$-valued random variables $X'$ and $Y'$ on $\mathcal{P}'$ such that $X' \sim F$, $Y' \sim G$, and $\E(Y'|X')=X'$ a.s. Since $\mathcal{P}$ is rich relative to $Y$, there exist $[0,1]$-valued random variables $X''$ and $Y''$ on $\mathcal{P}$ such that $(X'',Y'')$ has the same joint distribution as $(X',Y')$.%
		\footnote{Richness implies the existence of one uniformly distributed random variable on $\mathcal{P}$, and this implies that there are two independent random variables $U$ and $V$ on $\mathcal{P}$, each of which is uniformly distributed \parencite[e.g. Lemma~3.21 in][]{Kallenberg2002}. Let $(H_x)_{x \in [0,1]}$ be a regular conditional distribution function of $Y'$ given $X'$ (intuitively, $H_x$ is the CDF of $Y'$ conditional on $X'=x$). For any distribution $H$, write $H^{-1}(q) \coloneqq \inf\left\{ z \in [0,1] : H(z) \geq q \right\}$. Define $X'' \coloneqq F^{-1}(U)$ and $Y'' \coloneqq H_{X''}^{-1}(V)$. ($Y''$ is measurable because $(x,q) \mapsto H_x^{-1}(q)$ is jointly Borel-measurable, being Borel-measurable in its first argument and left-continuous in its second argument.)}
	Then, since $\mathcal{P}$ is rich relative to $Y$, Kallenberg's transfer theorem implies that there exists a random variable $X$ on $\mathcal{P}$ such that $(X,Y)$ has the same joint distribution as $(X'',Y'')$, hence the same joint distribution as $(X',Y')$, which implies in particular that $X \sim F$ and $\E(Y|X)=X$ a.s.
\end{proof}

\section{Feasible posterior-mean distributions}
\label{sec:bayes}

There is a fixed probability space $\mathcal{P} = (\Omega,\mathcal{F},\mu)$. There is an uncertain state of the world, formally a $[0,1]$-valued random variable $Y$ on $\mathcal{P}$. We assume that $\mathcal{P}$ is rich relative to $Y$, and we write $G$ for the distribution of $Y$.

By a \emph{signal,} we mean a(ny) $[0,1]$-valued random variable on $\mathcal{P}$. Because $\mathcal{P}$ is rich relative to $Y$, this straightforward definition is equivalent to the more elaborate definition à la \textcite{Blackwell1951,Blackwell1953}, which specifies the distribution of the signal conditional on the state.%
	\footnote{Namely: a signal is $(M,\pi)$, where $M$ is a non-empty compact metric space and $\pi$ is a Borel-measurable map $[0,1] \to \Delta(M)$, where $\Delta(M)$ is the set of all Borel probability measures on $M$, with the topology of weak convergence. The interpretation is that $M$ is a set of messages, and that $\pi(x) \in \Delta(M)$ is the distribution of messages sent if the state is $x \in [0,1]$. Without loss of generality, we restrict attention to signals $(M,\pi)$ with $M = [0,1]$.}
The role of richness is to ensure that for any imaginable distribution conditional on the state $Y$, there exists a random variable $X$ which has that conditional distribution. (Such richness provisos are ubiquitous in economics, but usually left implicit.)

Given a signal, each signal realisation induces a posterior belief via Bayes's rule, whose expectation we call the \emph{posterior mean.}
Each signal thus induces a random posterior mean, with some distribution.
Call a distribution \emph{feasible given $Y$} if and only if it is the posterior-mean distribution induced by some signal.

The previous paragraph is loose. Formally, `posterior belief' means \emph{a conditional probability}, and `posterior mean' means \emph{a conditional expectation,} and these entities are unique only up to null sets (they are not pinned down signal realisation by signal realisation).

\begin{corollary}
	\label{corollary:feas}
	Fix a probability space $\mathcal{P}$, and let the state of the world be a $[0,1]$-valued random variable $Y$. Assume that $\mathcal{P}$ is rich relative to $Y$, and let $G$ denote the distribution of $Y$. Then, for a distribution $F$, the following are equivalent:

	\begin{itemize}
	
		\item $F$ is feasible given $Y$.

		\item $F$ belongs to $\mpc{G}$.
	
	\end{itemize}
\end{corollary}

The proof is a simple application of \Cref{theorem:kbhlp} (not of \hyperref[theorem:strassen]{Strassen's theorem}):

\begin{proof}
	Suppose that $F$ is feasible given $Y$. Then there exists a signal $Z$ such that $\E(Y|Z) \sim F$. Consider the $[0,1]$-valued random variable $X \coloneqq \E(Y|Z)$; it satisfies $X \sim F$, and furthermore $\E(Y|X) = \E(\E(Y|Z)|X) = \E(X|X) = X$ a.s. by the law of iterated expectations. It follows by \Cref{theorem:kbhlp} that $F$ belongs to $\mpc{G}$.

	Suppose that $F$ belongs to $\mpc{G}$. Then by \Cref{theorem:kbhlp}, there exists a $[0,1]$-valued random variable $X$ on $\mathcal{P}$ such that $X \sim F$ and $\E(Y|X)=X$ a.s. Consider the signal $Z \coloneqq X$, and let $M \coloneqq \E(Y|Z)$ be the random posterior mean induced by this signal. We have $M = \E(Y|X) = X$ a.s., so $M \sim F$, showing that $F$ is feasible given $Y$.
\end{proof}

Obviously \hyperref[theorem:strassen]{Strassen's theorem} cannot be substituted for \Cref{theorem:kbhlp} in this proof: we are after a random variable $X$ such that $\E(Y|X)=X$ a.s., but \hyperref[theorem:strassen]{Strassen's theorem} only delivers $\E(Y'|X)=X$ a.s. for some other random variable $Y'$ (in general distinct from $Y$, albeit with the same distribution).

\section{What if only $G$ is given?}
\label{sec:G_only}

It is common to specify the environment more tersely as `there is an uncertain state of the world, with distribution $G$', without explicit reference to the random variable $Y$ (or the underlying probability space). This environment is under-specified. We shall argue that this under-specification is harmless, but that \hyperref[theorem:strassen]{Strassen's theorem} remains beside the point.

Suppose that we \emph{do} apply \hyperref[theorem:strassen]{Strassen's theorem} to `characterise' feasible posterior-mean distributions. That yields the following result.

\begin{corollary}
	\label{corollary:feas_strassen}
	Let $G$ be the distribution of the state of the world. Then, for a distribution $F$, the following are equivalent:

	\begin{itemize}
	
		\item There exist a probability space $\mathcal{P}$ and a $[0,1]$-valued random variable $Y$ on $\mathcal{P}$ such that $Y \sim G$ and $F$ is feasible given $Y$.

		\item $F$ belongs to $\mpc{G}$.
	
	\end{itemize}
\end{corollary}

In other words, applying Strassen's theorem answers the following question: `what posterior-mean distributions can we induce (via Bayes's rule) if we can choose both the signal $Z$ \emph{and} the state of the world $Y$ (subject to the constraint $Y \sim G$)?' This is not an economically meaningful question, because in an economic model, the state of the world is exogenously given.

The fundamental issue is that the model is under-specified ($Y$ and the probability space are not given: only $G$ is). `Applying Strassen's theorem' amounts to interpreting this under-specification as granting us licence to choose both the signal and the state (and the probability space), which is not economically meaningful. The more reasonable interpretation of this under-specified model is that there \emph{is} a fixed exogenous probability space and state of the world $Y$ (a random variable whose distribution is $G$), even if this was not explicitly mentioned. Then feasible posterior-mean distributions are characterised by \Cref{corollary:feas} in the previous section.

The fact that this model is under-specified is ultimately harmless, because one way or another, the answer to `which posterior-mean distributions are feasible?' is `those in $\mpc{G}$'.

There is one upshot, though: it does \emph{not} make economic sense to attribute the `$\mpc{G}$' characterisation of feasible posterior-mean distributions to \hyperref[theorem:strassen]{Strassen's theorem}, which is relevant only if the state of the world is chosen (subject to a distribution constraint). Instead, the relevant result is \Cref{theorem:kbhlp}.

\section{Postscript on stochastic differential equations}
\label{sec:sde}

The distinction between \hyperref[theorem:strassen]{Strassen's theorem} and \Cref{theorem:kbhlp} is somewhat analogous to the distinction between weak and strong solutions of stochastic differential equations (SDEs).%
	\footnote{See e.g. \textcite{Oksendal1998} for an introduction to SDEs. My perspective on weak/strong solutions is inspired by Bruno Strulovici's thoughts, some of which can be found in \textcite{StruloviciSzydlowski2015}. Obviously any errors or excesses are mine.}
A typical SDE is
\begin{equation}
	\dd X_t = \nu(X_t,t) \dd t + \sigma(X_t,t) \dd B_t , \quad X_0=k ,
	\tag{$\partial$}
	\label{sde}
\end{equation}
where $\nu$ and $\sigma$ are given functions, $B$ is a given Brownian motion, and $k$ is a given constant. The interpretation is that \eqref{sde} describes how the endogenous stochastic process $X$ is driven by the exogenous stochastic forcing term $\dd B_t$. In the background, there is a given filtered probability space
\begin{equation*}
	\mathcal{P}=(\Omega,\mathcal{F},(\mathcal{F}_t)_{t \in \R_+},\mu)
\end{equation*}
which carries the process $B$. We assume that $\mathcal{P}$ satisfies the so-called `usual conditions', and that $B$ is an $(\mathcal{F}_t)_{t \in \R_+}$-Brownian motion.

A \emph{strong solution} of \eqref{sde} is a process $X$ on $\mathcal{P}$ which is continuous, suitably adapted, and suitably integrable, and for which almost surely, \eqref{sde} holds in the Itô sense for every $t$. A \emph{weak solution} of \eqref{sde} is a triplet $(\mathcal{P}',B',X')$, where
\begin{equation*}
	\mathcal{P}'=(\Omega',\mathcal{F}',(\mathcal{F}'_t)_{t \in \R_+},\mu')
\end{equation*}
is a filtered probability space satisfying the usual conditions, $B'$ is an $(\mathcal{F}'_t)_{t \in \R_+}$-Brownian motion, and $X'$ is a process on $\mathcal{P}'$ which is continuous, suitably adapted, and suitably integrable, and for which almost surely,
\begin{equation*}
	\dd X'_t = \nu(X'_t,t) \dd t + \sigma(X'_t,t) \dd B'_t , \quad X'_0=k 
	\tag{$\partial'$}
	\label{sdetilde}
\end{equation*}
holds in the Itô sense for every $t$. So `really', a weak solution of \eqref{sde} is a solution of another SDE, \eqref{sdetilde}, which looks like \eqref{sde} but is defined on some other filtered probability space, $\mathcal{P}'$, and has a different exogenous stochastic forcing term, $B'$ (albeit with the same law: it is a Brownian motion).

Weak solutions are used e.g. in physics, but also in economics and finance. They are convenient because they have better existence properties than strong solutions. But in an economic model, weak solutions make little sense: the forcing term $B$ is exogenously given, so replacing it by some other term $B'$ is not economically meaningful.



\printbibliography[heading=bibintoc]

\end{document}